%% file: main.tex
\documentclass[runningheads]{llncs}
\usepackage{graphicx}
\usepackage{pdfpages}
\usepackage{caption}
\usepackage{subcaption}

\usepackage{tikz}
\tikzset{>=latex} 
\usetikzlibrary{calc}
\usetikzlibrary{arrows.meta}
\usetikzlibrary{backgrounds}
\usetikzlibrary{patterns}
\usepackage{pgfplots}
\usepackage{enumitem}

\usepgfplotslibrary{fillbetween}
\colorlet{veccol}{green!45!black}
\colorlet{myred}{red!90!black}
\colorlet{myblue}{blue!90!black}
\colorlet{mypurple}{blue!50!red!80!black!80}
\tikzstyle{vector}=[->,very thick,veccol]
\usetikzlibrary{arrows.meta}
\tikzstyle{thin arrow}=[dashed,thin,-{Latex[length=4,width=3]}]

\tikzset{
  treenode/.style = {shape=rectangle, rounded corners,
                     align=center,
                     fill=gray!20, outer sep=2pt},
  root/.style     = {treenode  },
  accept/.style      = {treenode, fill=SpringGreen!90, inner sep=7pt, draw},
    discard/.style      = {treenode, fill=Red!60, inner sep=7pt, draw},
  dummy/.style    = {circle,fill=MidnightBlue!70, inner sep=5pt, outer sep=2pt, draw}
}
\usetikzlibrary{decorations.text}

\usepackage{booktabs}
\usepackage{amsmath}
\usepackage{amsfonts} 
\usepackage[square,sort&compress,comma,numbers]{natbib}
\usepackage{hyperref,cleveref}
\usepackage{xspace}
\usepackage{subcaption}

\usepackage{color}
\definecolor{asparagus}{rgb}{0.53, 0.66, 0.42}
\definecolor{bananamania}{rgb}{0.98, 0.91, 0.71}
\definecolor{cadmiumorange}{rgb}{0.93, 0.53, 0.18}
\definecolor{cerisepink}{rgb}{0.93, 0.23, 0.51}
\definecolor{cambridgeblue}{rgb}{0.64, 0.76, 0.68}
\definecolor{brightube}{rgb}{0.82, 0.62, 0.91}
\usepackage{pgfplots}

\def\framework{KGUF\xspace}

\makeatletter
\newcommand\feature[1]{%
  \@tempcnta=0
  \langle
  \@for\@ii:=#1\do{%
    \@insertbreakingcomma
    \textit{\@ii}
  }%
  \rangle
}
\def\@insertbreakingcomma{%
  \ifnum \@tempcnta = 0 \else, \linebreak[1] \fi
  \advance\@tempcnta\@ne
}
\makeatother

\begin{document}
\title{KGUF: Simple Knowledge-aware\\ Graph-based Recommender with\\ User-based Semantic Features Filtering}
\titlerunning{KGUF: Simple Knowledge-aware Graph-based Recommender}
%
\author{Salvatore Bufi\inst{1} \and
Alberto Carlo Maria Mancino\inst{1} \and
Antonio Ferrara\inst{1} \and \\
Daniele Malitesta\inst{2}\thanks{Work done while at Politecnico di Bari.} \and
Tommaso Di Noia\inst{1} \and
Eugenio Di Sciascio\inst{1}}%

\institute{Politecnico di Bari,
\email{name.surname@poliba.it}\\
\and
Université Paris-Saclay, CentraleSupélec, Inria,
\email{name.surname@centralesupelec.fr}}

\authorrunning{Bufi, Mancino, Ferrara, et al.}

\maketitle              
\begin{abstract}

The recent integration of Graph Neural Networks (GNNs) into recommendation has led to a novel family of Collaborative Filtering (CF) approaches, namely Graph Collaborative Filtering (GCF). Following the same GNNs wave, recommender systems exploiting Knowledge Graphs (KGs) have also been successfully empowered by the GCF rationale to combine the representational power of GNNs with the semantics conveyed by KGs, giving rise to Knowledge-aware Graph Collaborative Filtering (KGCF), which use KGs to mine hidden user intents. 
Nevertheless, empirical evidence suggests that computing and combining user-level intent might not always be necessary, as simpler approaches can yield comparable or superior results while keeping explicit semantic features. Under this perspective, user historical preferences become essential to refine the KG and retain the most discriminating features, thus leading to concise item representation. Driven by the assumptions above, we propose \framework, a KGCF model that learns latent representations of semantic features in the KG to better define the item profile. By leveraging user profiles through decision trees, \framework effectively retains only those features relevant to users. Results on three datasets justify \framework's rationale, as our approach is able to reach performance comparable or superior to SOTA methods while maintaining a simpler formalization.

\end{abstract}

\input{sections/introduction}

\input{sections/related_work}

\input{sections/background}

\input{sections/methodology}

\input{sections/experimental_settings}

\input{sections/results_discussion}

\input{sections/conclusion}
%
%
%
%
%
\bibliographystyle{splncs04}
\bibliography{bibliography}
\end{document}

%% file: sections/introduction.tex
\section{Introduction}
Recommender Systems find widespread application across diverse domains, owing to their proficiency in furnishing users with personalized items~\cite{DBLP:conf/recsys/CovingtonAS16,DBLP:conf/kdd/YingHCEHL18,DBLP:conf/sigir/Yuan0KZ20}. Typically, these systems rely on past user-item interactions to extrapolate new preferences and generate personalized suggestions~\cite{DBLP:conf/sigir/EbesuSF18,DBLP:conf/www/HeLZNHC17}.
Collaborative Filtering~\cite{DBLP:journals/computer/KorenBV09} (CF) methods adeptly extract similarity patterns from user-item interactions, constituting (to date) the most diffused and winning recommendation paradigm~\cite{DBLP:conf/uai/RendleFGS09,DBLP:conf/www/HeLZNHC17,DBLP:conf/cikm/MaoZWDDXH21}. Nevertheless, by design, the learned user and item profiles in traditional CF recommender systems cannot embed the information conveyed by the underlying multi-hop user-item relations, thus disregarding a rich source of potentially-useful information~\cite{DBLP:conf/sigir/Wang0WFC19,DBLP:conf/cikm/MaoZXLWH21}. Following the popular graph representation learning~\cite{DBLP:series/synthesis/2020Hamilton} wave, Graph Neural Networks (GNNs) have recently been applied to CF approaches, giving rise to a fresh family of recommender systems~\cite{DBLP:conf/sigir/0001DWLZ020,DBLP:conf/cikm/MaoZXLWH21,DBLP:conf/sigir/WuWF0CLX21,DBLP:journals/tkde/YuXCCHY24}, better known as Graph Collaborative Filtering (GCF). 
Despite the success of traditional CF and GCF recommendation systems, such approaches often overlook the nuances in user preferences related to specific item characteristics; indeed, this represents a missed opportunity for enhancing user personalization. As opposed to CF systems, content-based recommendation~\cite{DBLP:reference/sp/GemmisLMNS15} leverages the attributes of historically-favored products, proposing unseen items that align with user distinct tastes~\cite{DBLP:conf/www/HeLZNHC17}.
In these systems, Knowledge Graphs (KGs) have been shown to provide a meaningful source of structured information~\cite{DBLP:conf/recsys/AnelliNSFM21, DBLP:conf/sigir/Zou0MWQ0C22, DBLP:conf/sigir/0002ZCZG22, DBLP:conf/cikm/Zou0WM0FC22} used to craft features that enrich the item description, thanks to the mapping between items and semantic entities, and helps to tackle open challenges in recommendation, such as the graph sparsity.

As a recent trend, an increasing number of recommender system algorithms currently propose to combine the representational capabilities of GNNs with the rich information conveyed by KGs, resulting in a novel hybrid family of recommendation techniques which might be suitably named as Knowledge-aware Graph Collaborative Filtering~\cite{DBLP:conf/sigir/YangHXL22, DBLP:conf/sigir/Zou0MWQ0C22, DBLP:conf/cikm/HuangHC21,DBLP:conf/www/WangHWYL0C21} (KGCF). Most of KGCF methods model the user intent to enhance recommendation quality and interpretability, thus outperforming similar baselines techniques~\cite{DBLP:conf/kdd/Wang00LC19,DBLP:conf/www/WangHWYL0C21,DBLP:conf/recsys/MancinoFBMNS23}. However, such architectures often require convoluted aggregations of several ad-hoc methodologies and strategies to learn meaningful user and item profiles, making it hard to apply KGCF approaches in real-world scenarios.
In this regard, we believe that these architectures can be dramatically simplified while maintaining comparable or even superior performance. In particular, we consider to utilize the content information from KGs solely for describing item representations, refraining from any attempt to directly incorporate it into building user profiles, which can be instead indirectly learned by means of collaborative filtering techniques.


\begin{figure}[!t]
\centering
\begin{subfigure}[c]{.5\textwidth}
  \centering
  \includegraphics[width=.95\linewidth]{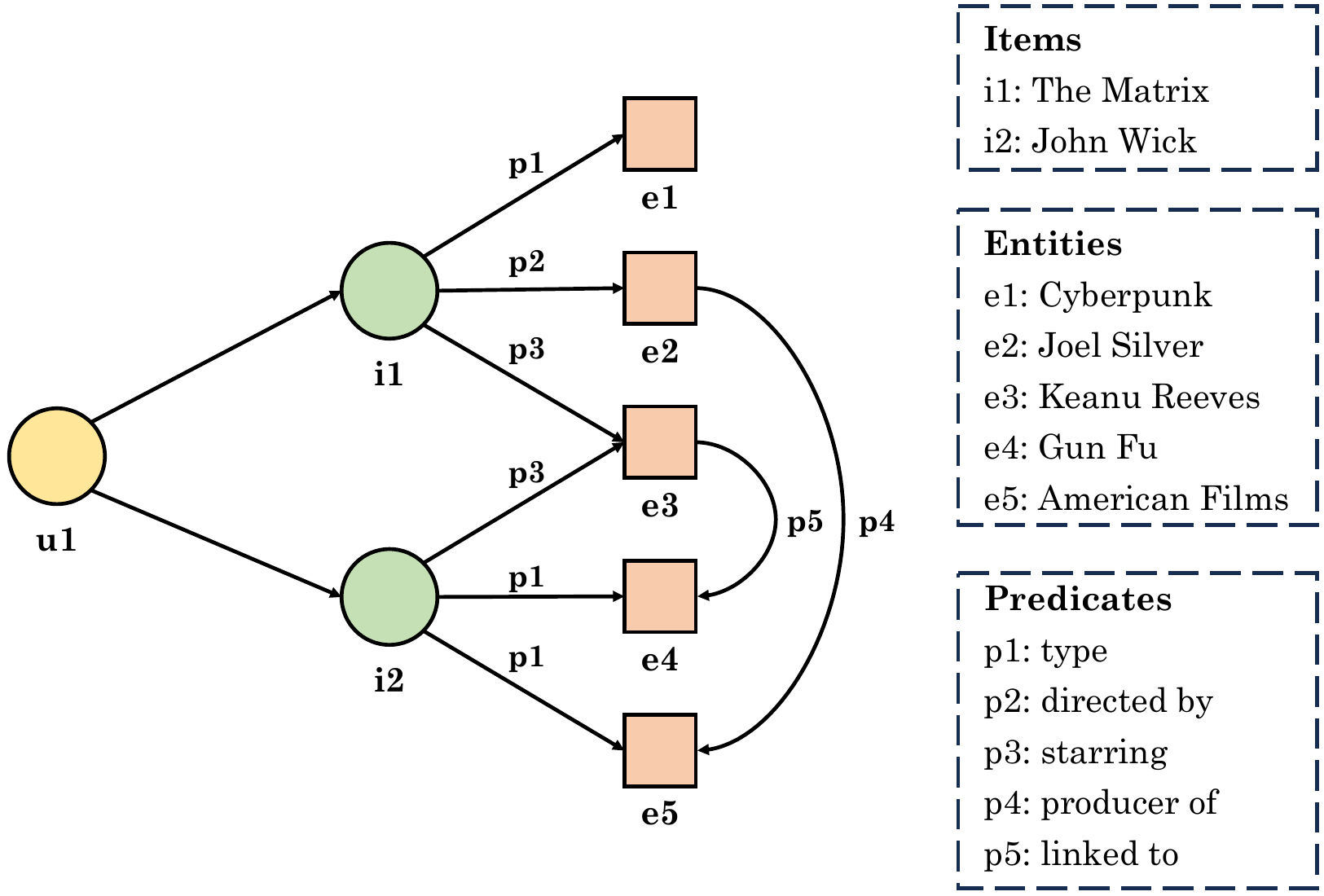}
  \caption{}
  \label{fig:sub1}
\end{subfigure}
\begin{subfigure}[c]{.45\textwidth}
  \centering
  \includegraphics[width=.95\linewidth]
  {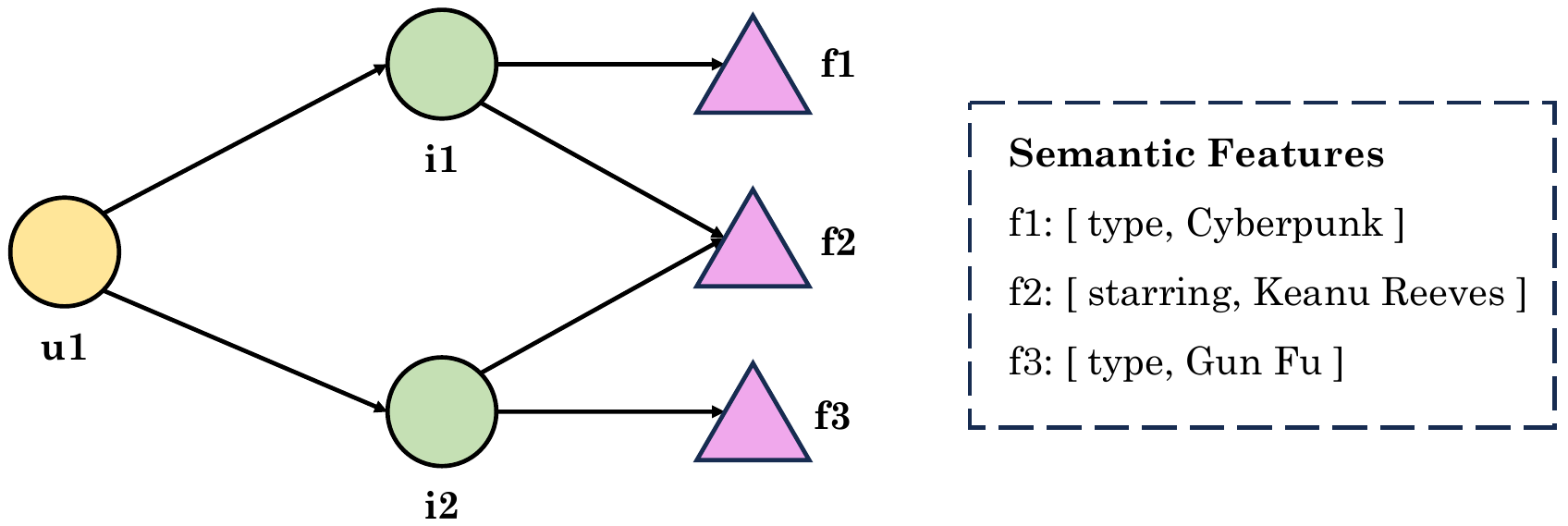}
  \caption{}
  \label{fig:sub2}
\end{subfigure}
\caption{An example of how \framework models the item representation. Figure \ref{fig:sub1} shows the entities in the knowledge graph linked to the items. In figure \ref{fig:sub2}, the most relevant semantic features selected by \framework are reported.}
\label{fig:test}
\end{figure}

Driven by these assumptions, this paper presents a novel KGCF recommendation algorithm named \textbf{\framework} (short for \textbf{K}nowledge \textbf{G}raph \textbf{U}ser-based \textbf{F}iltering) designed to efficiently harness side information from KGs. From a technical perspective, \framework learns user and item embeddings through linear propagation on the collaborative graph. In contrast to other approaches, our solution, sketched in \Cref{fig:test}, is able to integrate latent representations for semantic features right into the item representation. Moreover, sharing the same principle but replacing it with a simple mechanism, we propose substituting the intent modeling with a decision tree mechanism that selects the most meaningful semantic features in the KGs and filters out noisy features that poorly contribute to the item description but heavily impact the computational costs.
All in all, the key contributions of this paper are summarized as follows:
\begin{enumerate}[leftmargin=*]
    \item We propose a novel recommendation model, namely \framework, which effectively selects and integrates user-relevant semantic features during the graph learning phase to improve item representation. 
    \item We conduct extensive experiments over three well-known recommendation datasets to assess the efficacy of our framework and evaluate the contribution of collaborative and content signals on \framework.
    \item We delve into using decision trees for selecting semantic features relevant to users, investigating how the effects of negative sampling and the presence of constraints regarding the tree depth affect the recommendation performance.
\end{enumerate}

To foster the replicability of the experiments and results, we release the code for \framework and the baselines, along with the  datasets and configurations~\footnote{~\url{https://github.com/sisinflab/KGUF}}.

%% file: sections/related_work.tex
\section{Related work}
This section aims to survey and compare other approaches in recommendation literature leveraging KGs and graph representation learning strategies. \\

\noindent \textbf{KGs information in recommendation.}
 As a general trend, the incorporation of information coming from KGs into recommendation systems has consistently demonstrated improvements in model performance and interpretability 
across diverse domains~\cite{DBLP:conf/recsys/AnelliNSFM21, DBLP:conf/sigir/Zou0MWQ0C22, DBLP:conf/sigir/0002ZCZG22, DBLP:conf/cikm/Zou0WM0FC22}. Indeed, the utilization of KGs has been effective in boosting different recommendation tasks, including hybrid collaborative/content-based recommendation \cite{DBLP:conf/IEEEwisa/LiXTZT20,DBLP:conf/recsys/AnelliNSFM21}, where the KG is of support to the lack of collaborative information, eventually enhancing performance; explainable recommendation \cite{DBLP:series/ssw/47,DBLP:journals/tkde/AnelliNSRT22}, where the KG serves as a robust information source for delivering human-like explanations; and user modeling \cite{DBLP:journals/tois/WangZWZLXG19,DBLP:conf/semweb/AnelliNSRT19}, with the semantic description of items contributing to defining user profiles.
KGs also offer a solution to the cold-start problem by augmenting observed interactions with KG-reachable unobserved items~\cite{DBLP:conf/wsdm/TogashiOS21}. However, it is crucial to note that in some cases, improper extraction of information may potentially weaken the model capabilities \cite{DBLP:conf/icde/ChenYWBSK22}.\\

\noindent \textbf{Graph representation learning for recommendation.}
Recommender systems based on graph representation learning strategies have demonstrated their superior performance by mining high-order user-item relationships embedded within the graph structure~\cite{DBLP:conf/sigir/ShuaiZWSHWL22, DBLP:conf/sigir/TianXLYZ22, DBLP:conf/sigir/WuWF0CLX21, DBLP:conf/cikm/ShenWZSZLL21}.
Inspired by the graph convolutional network layer (GCN) \cite{DBLP:conf/iclr/KipfW17}, early approaches like PinSage \cite{DBLP:conf/kdd/YingHCEHL18} and NGCF \cite{DBLP:conf/sigir/Wang0WFC19} have adapted GCN to the user-item bipartite graph. Later research has explored simplifying the GCN layer, as seen in methods like LightGCN \cite{DBLP:conf/sigir/0001DWLZ020} and LR-GCCF \cite{DBLP:conf/aaai/ChenWHZW20}, which streamline model complexity by eliminating non-linearities and feature transformations in node embeddings. Additionally, UltraGCN \cite{DBLP:conf/cikm/MaoZXLWH21} proposes a mathematical proxy to the infinite message-passing layer, thus limiting the negative effects of over-smoothing.
Another avenue of research focuses on pruning noisy user-item interactions to discern meaningful preference patterns. Inspired by the graph attention network \cite{DBLP:conf/iclr/VelickovicCCRLB18} (GAT), methodologies like DGCF \cite{DBLP:conf/sigir/WangJZ0XC20} aim to identify the underlying intents of individual user-item interactions on a fine-grained level. Furthermore, works such as \cite{DBLP:conf/cikm/AnelliDNSFMP22} prove how user-item edges can incorporate multimodal content, such as user reviews, to enhance recommendation and address the challenge of over-smoothing. \\

\noindent \textbf{KGs information and graph representation learning for recommendation.} 
Recent literature \cite{DBLP:conf/sigir/YangHXL22, DBLP:conf/sigir/Zou0MWQ0C22, DBLP:conf/cikm/HuangHC21} has ventured beyond traditional collaborative filtering graph-based methods by incorporating information from KGs to enhance item descriptions. One example is KGAT \cite{DBLP:conf/kdd/Wang00LC19}, which explicitly models the KG in an end-to-end manner. It refines node embeddings using an attention mechanism to discern the importance of neighbors.
In a different approach, the authors of KGIN \cite{DBLP:conf/www/WangHWYL0C21} utilize the KG to define users' intents as weights, combining these weights with their relations. CKAN \cite{DBLP:conf/sigir/WangLTCL20} proposes a method that integrates collaborative signals with knowledge data. More recently, KGs have been employed to formulate first-order logical rules, combined with graph embeddings, enabling recommenders based on neuro-symbolic reasoning \cite{DBLP:conf/aiia/SpilloMGLS22}.

%% file: sections/background.tex
\section{Background}
In this section, we introduce the notion of KGs and the basics of decision trees used for content modelling, along with the problem we aim to address.

\subsection{Knowledge graphs for recommendation}
A knowledge graph stores structured information of real-world facts in the form of a heterogeneous graph which can be used to retrieve item attributes~\cite{DBLP:conf/www/0003W0HC19,DBLP:conf/kdd/GaoLWWL11}.
Formally, it is presented as a set of triples in the form $\sigma \xrightarrow{\rho} \omega$. Each triple indicates that the head entity $\sigma$ is connected to the tail entity $\omega$ through the relation $\rho$. 
Given a set of entities $\mathcal{V}$ and a set of relations $\mathcal{R}$, a knowledge graph $\mathcal{KG}$ can be represented as follows:
\begin{equation}
    \label{eq:feature-expl}
    \mathcal{KG} = \{  \sigma \xrightarrow{\rho} \omega  \;|\; \sigma, \omega \in \mathcal{V}, \rho \in \mathcal{R} \}.
\end{equation}
Mapping items in the catalog using knowledge graphs, i.e., $\mathcal{I} \subseteq \mathcal{V}$, enables the descriptions of items based on semantic features.

\subsection{Decision trees}
A decision tree is a hierarchical tree-like structure used in machine learning for classification tasks. It iteratively partitions the dataset into subsets depending on the values of input features, resulting in a tree structure in which each internal node represents a feature-based decision.The decision-making process is guided by criteria such as Gini impurity or information gain.

In practical terms, decision trees aim to identify variables whose values contribute the most to reducing another random variable's entropy, or uncertainty.
Mathematically, the entropy $H(V)$ of a random variable $V$ with $k$ possible values in ${v_1, ..., v_k}$ is formally defined as:
\begin{equation}
\label{eq:entropy}
    H(V) = - \sum_{i=1}^k \mathrm{Pr}(V = v_i) \log_2\mathrm{Pr}(V = v_i).
\end{equation}

If observing the value of another variable $X$ with $n$ possible values in ${x_1, ..., x_n}$ results in a reduction of the entropy of $V$, then $X$ is said to exhibit information gain over $V$:
\begin{equation}
\begin{gathered}
\label{eq:infogain}
IG(V, X) = H(V) - H(V|X), \\
H(V|X) = \sum_{i=1}^n P(X = x_i) H(V | X = x_i).
\end{gathered}
\end{equation}
Decision trees are one possible lightweight solution to filter out noisy semantic features in the KG and select the most relevant ones from the user perspective.


%% file: sections/methodology.tex
\section{Methodology}
In this section, we present \framework and its components. In particular, we delve into the semantic item modeling and its refinements based on user decisions. Then, we introduce our simple but effective knowledge-aware aggregation schema.

\subsection{Modeling items through knowledge graph}
Assuming that each item in the item set $\mathcal{I}$ of our recommendation problem is associated with an entity in $\mathcal{V}$, essentially being one of the entities within the knowledge graph ($\mathcal{KG}$), we can model the items in the catalog by leveraging the knowledge embedded in the KG. 
The rationale behind this modeling lies in the notion that each item can be comprehensively characterized by a distinct set of features, some of them influencing user decisions in varying ways. 

Given this premise, the KG can be explored to identify connections with other entities for each item, serving as its attributes in the form $\sigma \xrightarrow{\rho} \omega$, where $\rho$ and $\omega$ represent the predicate and object of the KG triple, respectively. 
As an example, if properly encoded in the reference knowledge graph, the movie \textit{Harry Potter} could be semantically described by the predicate-object pairs $\feature{director, Chris Columbus}$ and $\feature{genre, Fantasy}$. 
Formally, we utilize the set of semantic features to depict the item $i \in \mathcal{I}$ based on its first-order connectivity extracted from the $\mathcal{KG}$:
\begin{equation}
    \label{eq:feature-expl}
    \mathcal{F}_i = \{ \langle \rho, \omega\rangle \mid i \xrightarrow{\rho} \omega \in \mathcal{KG} \}.
\end{equation}

\subsection{Refining item features  with user decision}
When confronted with a novel item, users focus on a subset of its numerous features.
To capture this user behavior, \framework uses decision trees to conduct a behavioral analysis at the user level, leveraging her past ratings. 
The objective is to discern the most pertinent features within the knowledge graph that effectively describe items in the catalog while filtering out noisy ones.
This tailored approach ensures that the model accentuates features that hold significant influence across all users in the system.

Formally, let $\mathcal{I}_u^+ \subseteq \mathcal{I}$ denote the set of items positively rated by each user $u$ in the set $\mathcal{U}$ of our recommendation problem, and let $\mathcal{I}_u^- = \mathcal{I} \setminus \mathcal{I}_u^+$ be the set of items not enjoyed by user $u$.
Let $\mathcal{\tilde{I}}_u^- \subseteq \mathcal{I}_u^-$ be a selection of items randomly sampled from $\mathcal{I}_u^-$, such that $|\mathcal{\tilde{I}}_u^-| = |\mathcal{I}_u^+|$. Finally, $\mathcal{I}_u = \mathcal{I}_u^+ \cup \mathcal{\tilde{I}}_u^-$ is the set of items selected to depict the past behavior of the user.

For each user $u$, the set of semantic features $\mathcal{F}_u = \bigcup_{i \in \mathcal{I}_u} \mathcal{F}_i$ is used to compute the user decision tree $T_u$ that, accordingly to the definition of Information Gain~\cite{DBLP:conf/recsys/AnelliNSFM21}, is able to find the minimum subset of semantic features $\mathcal{F}_{u}^{T_u} \subseteq \mathcal{F}_u$ that can distinguish whether an item belongs to $\mathcal{I}_u^+$ or $\mathcal{\tilde{I}}_u^-$. 
Finally, each item $i$ is formally described by the subset of its semantic features that appear in at least one user decision tree:
\begin{equation}
    \mathcal{F}_i^* = \mathcal{F}_i \cap \bigcup_{u \in \mathcal{U}} \mathcal{F}_{u}^{T_u} 
\end{equation}

\subsection{Knowledge-aware aggregation}

Inspired by LightGCN~\cite{DBLP:conf/sigir/0001DWLZ020}, the embeddings of each user and item, denoted as $\mathbf{e}_u \in \mathbb{R}^d$ and $\mathbf{e}_i \in \mathbb{R}^d$ respectively, are acquired through the recursive aggregation of their multi-hop neighbors' representations via message propagation across $L$ convolutional layers. The resulting representations at various levels, $(\mathbf{e}_u^{(0)}, \dots, \mathbf{e}_u^{(L)})$ and $(\mathbf{e}_i^{(0)}, \dots, \mathbf{e}_i^{(L)})$, emerge from the integration of content information with collaborative signals originating from first-hop neighbor nodes and propagated from the preceding layer.

By expressing knowledge graph features through latent vectors $\mathbf{e}_f$, \framework adeptly represents items based on their explicit characteristics. 
This modeling approach furnishes an explicit content-based representation for each item, situating it within the latent feature space as an amalgamation of vectors delineating its distinct characteristics.
Formally:
\begin{gather}
    \begin{aligned}
            \mathbf{e}_{u}^{(l)} &= \sum_{i \in \mathcal{N}(u)} \frac{1}{\sqrt{|\mathcal{N}(u)|}\sqrt{|\mathcal{N}(i)|}} \mathbf{e}_i^{(l - 1)}, \\
            \mathbf{e}_{i}^{(l)}&=  \frac{\alpha}{|\mathcal{F}_i^*|}\sum_{f \in \mathcal{F}_{i}^*} \mathbf{e}_f + ( 1 - \alpha ) \sum_{u \in \mathcal{N}(i)} \frac{1}{\sqrt{|\mathcal{N}(u)|}\sqrt{|\mathcal{N}(i)|}} \mathbf{e}_u^{(l - 1)} ,         
    \end{aligned}
    \label{eq:aggregation}
\end{gather}
where $\alpha$ serves as a weight factor for balancing content and collaborative contributions in item representations.

It is worth noting that, in contrast to LightGCN, \framework introduces a knowledge-aware correction term in the item embeddings, which will affect also user representation through message passing schema. The collaborative and content signals collaboratively contribute to capturing long-range influences and leveraging the rich information derived from knowledge graphs. 
The use of decision trees to mediate content information guarantees that only relevant features engage in embedding modeling, ensuring a succinct and meaningful representation of users and items.

After completing $L$ propagation hops, the user and item embeddings are derived as follows:
    \begin{gather}
      \begin{aligned}
        \mathbf{e}_u &= \sum_{l = 0}^{L} \frac{1}{1 + l} \mathbf{e}^{(l)}_u, \quad \mathbf{e}_i &= \sum_{l = 0}^{L} \frac{1}{1 + l} \mathbf{e}^{(l)}_i. \\
      \end{aligned}\label{eq:final_emb}
    \end{gather}
The introduction of the scaling factor $1/(1 + l)$ serves to mitigate the oversmoothing problem that may arise with an increased number of explored hops~\cite{DBLP:conf/sigir/0001DWLZ020}.


\subsection{Model learning}
The training stage involves learning the values of user, item, and feature embeddings to align user-item estimated interactions with the actual preferences of users.
Given the embeddings $\mathbf{e}_u$ and $\mathbf{e}_i$, we employ the dot product of user and item representations as the prediction, expressed as:
\begin{equation}
    \hat{r}_{ui} = \mathbf{e}^\top_u \mathbf{e}_i.
    \label{eq:final_pred}
\end{equation}

To optimize the model parameters, we utilize the pairwise optimization algorithm  Bayesian Personalized Ranking (BPR). This algorithm operates on the assumption that a user $u$ favors a consumed item $i^+$ over a non-consumed item $i^-$. It optimizes the model by maximizing, for each pair of $i^+$ and $i^-$, a function based on the difference $r_{ui^+} - r_{ui^-}$. Specifically, by constructing a training set $\mathcal{T} = \{ (u, i^+, i^-) \mid (u, i^+) \in \mathcal{R}, (u, i^-) \notin \mathcal{R}, i^- \in \mathcal{I} \}$, BPR aims to optimize the following loss, where $\sigma$ represents the sigmoid function:
\begin{equation}
    \mathcal{L}_{\mathrm{BPR}} = \sum_{(u, i^+, i^-) \in \mathcal{T}} -\ln \sigma(\hat{r}_{ui^+} - \hat{r}_{ui^-}),
\end{equation}
Finally, we use a joint learning approach to include the L2-regularization loss as follows:
\begin{equation}
        \mathcal{L} = \mathcal{L}_{\mathrm{BPR}} + \lambda \lVert \Theta \rVert^2_2,
\end{equation}
here $\Theta$ encompass all the learnable parameter, specifically $\Theta = \{\mathbf{e}_u, \mathbf{e}_i, \mathbf{e}_f \mid u \in \mathcal{U}, i \in \mathcal{I}, f \in \bigcup_{i\in\mathcal{I} }\mathcal{F}_i^* \}$, while $\lambda$ controls the regularization term.


%% file: sections/experimental_settings.tex
\section{Experimental settings}
This section delineates the experimental configuration employed to evaluate the recommendation performance of \framework.

\subsection{Datasets and Knowledge Graph}
We compare the performance of \framework with state-of-the-art solutions on three well-known recommendation datasets:  \textit{MovieLens 1M}\footnote{MovieLens 1M Dataset: \url{https://grouplens.org/datasets/movielens/1m/}}, \textit{Yahoo! Movies} \footnote{Yahoo! Movies User Ratings and Descriptive Content Information, v.1.0: \url{https://webscope.sandbox.yahoo.com/catalog.php?datatype=r}}, and \textit{Facebook Books}\footnote{Facebook Books (Linked Open Data challenge co-located with ESWC 2015): \url{https://2015.eswc-conferences.org/important-dates/call-RecSys.html}}.
MovieLens 1M and Yahoo! Movies are two datasets in the movie recommendation domain that collect explicit ratings (ranging from 1 to 5) from 6,040 and 4000 users, respectively, over catalogs of 3,706 and 2,626 items. Overall, they report 1,000,209 and 69,846 ratings, respectively.
\textit{Facebook Books} is a dataset that collects 18,978 implicit user feedback in the book recommendation domain. Since its ratings refer to 4,000 users and 2,626 items, it is characterized by greater sparsity.
All the datasets are evaluated in an implicit feedback scenario: explicit ratings are binarized by applying a threshold value of 3. Only ratings greater or equal to the threshold value are recorded as implicit positive feedback.
In order to reduce the noisy contribution of users and items with too few occurrences, we apply an iterative 5-core strategy, dropping users and items with less than five appearances.
Through explicit linkings, the items in the dataset catalogs are directly linked to their semantic representation in \textit{DBpedia}, a well-recognized public knowledge graph.
For a fair comparison, we filter the noisy item-linked entities following, for each model, the filtering approach reported by the original authors.
\subsection{Baselines}

We empirically validate the effectiveness of \framework by comparing its performance in terms of recommendation accuracy with different baselines from five related categories described in the following.
\begin{description}[leftmargin=1em]
\item[Unpersonalized] We test Random and Most Popular recommendations as a basis for discerning significance and behaviors influenced by popularity.
\item[Collaborative Filtering] BPR-MF~\cite{DBLP:conf/uai/RendleFGS09} and Item-kNN~\cite{DBLP:journals/tkdd/Koren10} are two widely recognized collaborative approaches. The former follows a pair-wise learning approach as \framework, and the latter exploits similarities between items by computing the nearest neighbors.
\item[Knowledge-Aware]
Among these models, we selected KaHFM~\cite{DBLP:conf/semweb/AnelliNSRT19} and  KGFlex~\cite{DBLP:conf/recsys/AnelliNSFM21, 10.1145/3588901}, since they share with \framework the feature representation but differ in the way they exploit it. KaHFM uses knowledge for initializing latent factors in factorization machines. KGFlex, instead, models the user-item interaction by means of a sparse collection of the most relevant semantic features.
\item[Graph-Based]
LightGCN~\cite{DBLP:conf/sigir/0001DWLZ020} rethinks the classic GCN aggregation schema by discarding feature transformation and non-linear activation functions. It results in linear aggregations with improved performance.
DGFC~\cite{DBLP:conf/sigir/WangJZ0XC20} introduces a graph disentangling strategy for intercepting hidden user intents to model intent-dependent user-item interactions.
\item[Graph-based with Knowledge]
KGIN~\cite{DBLP:conf/www/WangHWYL0C21} exploits the KG for computing users' intents as a weighted combination of semantic relations. User-item connections are interpreted by means of a limited set of relevant intents.
KGTORe~\cite{DBLP:conf/recsys/MancinoFBMNS23} mines the user motivations by looking at the discriminant semantic features in her history. These explicit features are then used to enrich the user and the item node representation. 
\end{description}

\subsection{Evaluation protocol and reproducibility}

For the sake of reproducibility, we provide our code, data, and comprehensive instructions to enable the replication of all stages in our experimental procedures\footnote{\url{https://github.com/sisinflab/KGUF}}. 
This paper follows the same evaluation protocol and experimental setting of~\cite{DBLP:conf/recsys/MancinoFBMNS23}. 
Moreover, in order to compare \framework with the same results, we adopt the reproducibility framework Elliot~\cite{DBLP:conf/sigir/AnelliBFMMPDN21}.
Each dataset is split into train, test, and validation sets by retaining 72\%, 20\%, and 8\% of the original user ratings, respectively \cite{DBLP:reference/sp/GunawardanaS15}, and adopting the \textit{all unrated protocol}~\cite{DBLP:conf/recsys/Steck13}. 


Coherently with \cite{DBLP:conf/recsys/MancinoFBMNS23}, the embedding size is fixed to 64, and each vector is randomly initialized using a Xavier initializer~\cite{DBLP:journals/jmlr/GlorotB10}. Then, we train \framework on 20 different combinations of hyperparameters explored with a Bayesian optimization algorithm and apply early-stopping with a 5-epoch patience. 
We validate the our model on nDCG@10, since this metric enhances the capability of predicting top-ranking items. Moreover, the batch size is adapted proportionally to the dataset size: 64, 256, and 2048, respectively, for \textit{Facebook Books}, \textit{Yahoo! Movies}, and \textit{Movielens 1M}. 

%% file: sections/results_discussion.tex
\section{Results and discussion}
This section examines the performance of \framework when compared to state-of-the-art baselines. 
Additionally, we present hyperparameter and ablation studies conducted to evaluate the effectiveness of the proposed approach.
Specifically, this section delves into experimental analyses aimed at answering the following research questions:
\begin{itemize}[leftmargin=*]
    \item \textbf{RQ1} \textit{Overall performance:} How does \framework performance compare to that of SOTA recommenders? Additionally, what is the impact of the proposed knowledge-aware aggregation schema on the overall performance?
    \item \textbf{RQ2} \textit{Decision Tree Construction:} To what extent does the construction of trees impact the performance of the proposed framework? 
\end{itemize}

\subsection{RQ1: Overall performance}
\input{tables/accuracy_noprec}

Firstly, we compare the recommendation accuracy of all the methods on three different datasets and report the results in Table \ref{table:accuracy}. We denote the highest-performing result by boldfacing it and underscore the second-best result.

We observe that \framework shows performance comparable or superior to all baselines over the three datasets.
This improvement can be attributed to two key factors.
Firstly, by employing user-based filtering, \framework selectively learns the most pertinent characteristics for describing items in the catalog. In other words, it effectively filters knowledge information that proves genuinely valuable for the recommendation task.

Secondly, the linear aggregation schema adeptly combines collaborative and knowledge signals, thus improving both item and user representations.
Another observation is that GNN-based techniques (LightGCN, DGCF, KGIN, KGTORe) consistently outperform non-graph-based ones, underscoring the efficacy of these architectures in leveraging long-term user-item relationships. The superior performance of knowledge-aware GNN-based methods is particularly noteworthy, highlighting how graph models effectively incorporate and exploit knowledge graphs.

Finally, we observe that not all knowledge-aware models (KGFlex, kaHFM) outperform pure collaborative approaches, emphasizing how merely utilizing knowledge-aware information may not necessarily lead to improved performance.

\input{tables/ablation_knowledge}

To delve deeper into assessing the efficacy of the proposed aggregation schema, we analyze how varying the weight assigned to the knowledge-aware component within the \framework impacts the outcome on the dataset \textit{Yahoo! Movies}.
The results in Table \ref{tab:alpha1} involve varying the weight of $\alpha$ in Eq. \ref{eq:aggregation} across a spectrum from $0.2$ to $0.8$, i.e., increasing the contribution of knowledge on the collaborative information.
The results show that searching for the best balance between content and collaborative contributions is necessary, which in this case is 40\% content and 60\% collaborative. However, it is even more interesting to question the collaborative and the knowledge signals' separate impact when switched on/off. Performance in Table \ref{tab:alpha2} makes it clear that removing the knowledge information affects the model capabilities harder than removing the collaborative one. This shows how \framework benefits from semantic information and how combining the two signals gives superior performance to considering only one.
\subsection{RQ2: Decision Tree Construction}
\label{tree_construction}


In this study, we explore how the tree construction strategies can impact the overall performance of \framework, assessing how the negative sampling augmentation for tree building and the maximum depth of the trees contribute to the overall recommendation quality. \\

\noindent \textbf{Effect of negative sampling.}
We investigate how \framework behaves when we enhance the negative sampling employed in constructing decision trees by introducing a number of negative items equal to $\eta$ times the number of positive items, such that $|\mathcal{\tilde{I}}_u^-| = \eta\; |\mathcal{I}_u^+|$.
The results depicted in Figure \ref{fig:npr_ablation} indicate an improvement in accuracy performance with respect to the target metric as $\eta$ increases. We attribute this trend to the higher values of $\eta$ mitigating the influence of randomly selecting negative items, consequently reducing noise impact. However, it is important to note that increasing the value of $\eta$ does not guarantee a constant improvement in performance. Therefore, there is no ideal value for $\eta$, and this parameter must be explored individually for each dataset. \\

\noindent \textbf{Effect of tree depth.}
The maximum depth of a decision tree, representing the length of the longest path from the tree root to a leaf, directly influences the selection of features. We investigate how imposing constraints on the tree depth may impact the recommendation performance.
To address this, we trained the model with diverse maximum tree depths—1, 2, 5, 10, 15, 20, and without a limit ($\infty$). We evaluated accuracy performance repeating the experiment five times and presenting the average, minimum, and maximum values of the target metric in Figure \ref{fig:depth_abl}.
While unlimited depth yields satisfactory results, it is noteworthy that exploring the entire tree incurs higher computational and memory costs. Excessive features can lead to an overly generalized model, and comparable or even superior performance can be achieved without traversing the entire tree, as it is evident at depth 10, where we witness the least variance around the average performance.

\input{images/ablations}

%% file: tables/accuracy_noprec.tex
\begin{table*}[!t]
\small
    \caption{Accuracy performance comparison of \framework with the selected baselines. All the models are grouped into their respective categories.}
    \label{table:accuracy}
    \centerline{
\begin{tabular}{llllllllll}
\toprule
& \multicolumn{3}{c}{\textbf{Facebook Books}} & \multicolumn{3}{c}{\textbf{Yahoo! Movies}} & \multicolumn{3}{c}{\textbf{MovieLens 1M}} \\
\cmidrule(r){2-4} \cmidrule(lr){5-7} \cmidrule(l){8-10}
 & \textbf{nDCG} & \textbf{HR} & \textbf{Recall} & \textbf{nDCG} & \textbf{HR} & \textbf{Recall} & \textbf{nDCG} & \textbf{HR} & \textbf{Recall} \\
\cmidrule(r){2-4} \cmidrule(lr){5-7} \cmidrule(l){8-10}
\textbf{Random} & 0.0074 & 0.0256 & 0.0110 & 0.0065 & 0.0342 & 0.0090 & 0.0096 & 0.0836 & 0.0036 \\
\textbf{MostPop} & 0.0787 & 0.2535 & 0.1117 & 0.1185 & 0.3238 & 0.1501 & 0.1730 & 0.6537 & 0.0773 \\
\midrule
\textbf{BPR-MF} & 0.0945 & 0.2973 & 0.1358 & 0.2303 & 0.5228 & 0.2843 & 0.3100 & 0.8811 & 0.1627 \\
\textbf{Item-kNN} & 0.1039 & 0.3353 & 0.1556 & 0.2258 & 0.5033 & 0.2658 & 0.2932 & 0.8504 & 0.1490 \\
\midrule
\textbf{KGFlex} & 0.0712 & 0.2235 & 0.1001 & 0.1758 & 0.4360 & 0.2022 & 0.0111 & 0.1058 & 0.0053 \\
\textbf{kaHFM} & 0.0843 & 0.2776 & 0.1221 & 0.2085 & 0.5064 & 0.2621 & 0.2992 & 0.8654 & 0.1526 \\
\midrule
\textbf{LightGCN} & 0.0969 & 0.3170 & 0.1435 & 0.2230 & 0.5085 & 0.2712 & 0.3081 & 0.8813 & 0.1614 \\
\textbf{DGCF} & 0.1001 & 0.3192 & 0.1444 & 0.2356 & 0.5308 & 0.2848 & 0.3124 & 0.8819 & 0.1637 \\
\midrule
\textbf{KGIN} & 0.1023 & 0.3112 & 0.1448 & 0.2368 & 0.5500 & \underline{0.2971} & 0.3082 & 0.8854 & 0.1638 \\
\textbf{KGTORe} & \textbf{0.1188} & \textbf{0.3601} & \textbf{0.1698} & \underline{0.2471} & \underline{0.5566} & 0.2944 & \underline{0.3212} & \underline{0.8957} & \underline{0.1710} \\
\midrule
\textbf{\framework} & \underline{0.1156} & \underline{0.3594} & \underline{0.1697} & \textbf{0.2561} & \textbf{0.5685} & \textbf{0.3058} & \textbf{0.3277} & \textbf{0.8998} & \textbf{0.1755} \\
\bottomrule
\end{tabular}
}
\vspace{-1em}
\end{table*}

%% file: tables/ablation_knowledge.tex
\begin{table}[!t]
\caption{Performance of \framework regarding nDCG@10 when varying the knowledge contribution. Table \ref{tab:alpha1} reports the impact of the parameter $\alpha$, which weights the knowledge contribution. Table \ref{tab:alpha2} shows the impact of considering the knowledge or the collaborative signals only.}
\centering
\begin{subtable}{.45\textwidth}
    \caption{}
    \label{tab:alpha1}
    \centering
    \begin{tabular}{lr}
    \toprule
        \textbf{Configuration} & \textbf{nDCG@10} \\
        \midrule
        $\framework_{\alpha=.2}$ &  0.25255\\
        $\framework_{\alpha=.4}$ &  \textbf{0.25640}\\
        $\framework_{\alpha=.6}$ &  0.25233\\
        $\framework_{\alpha=.8}$ &  0.25322\\
    \bottomrule
    \end{tabular}
\end{subtable}
\begin{subtable}{.45\textwidth}
    \caption{}
    \label{tab:alpha2}
    \centering
    \begin{tabular}{lr}
    \toprule
        \textbf{Setting} & \textbf{nDCG@10} \\
        \midrule
        $\framework_{\mathrm{w/o\_knowledge}}$ &  0.23031\\
        $\framework_{\mathrm{w/o\_collaborative}}$ &  0.25074\\
        $\framework_{\alpha=.4}$ &  \textbf{0.25640} \\

    \bottomrule
    \end{tabular}
\end{subtable}
\vspace{-1em}
\label{tab:ablation}
\end{table}


%% file: images/ablations.tex
\begin{figure}[!t]
\centering
\begin{subfigure}[t]{.44\textwidth}
    \begin{tikzpicture}
        \begin{axis}[
            width=\linewidth,
            height=5.5cm,
            xlabel={$\eta$},
            ylabel={nDCG@10},
            ymin=0.235,
            ymax=0.262,
            ybar,
            xtick align=inside,
            bar width=10,
            grid=major,
            symbolic x coords={1, 2, 5, 10, 20}, 
            xtick=data, 
            xlabel near ticks,
            ylabel near ticks
        ]
        \addplot table [col sep=tab, x index=0, y index=1] {images/csv/yahoo_npr_seed.tsv};
        \end{axis}
    \end{tikzpicture}
    \caption{}
    \label{fig:npr_ablation}
\end{subfigure}
\begin{subfigure}[t]{.54\textwidth}
\begin{tikzpicture}
    \begin{axis}[
    height=5.5cm,
    width=\linewidth,
    xlabel={max depth},
    ylabel={nDCG@10},
    ymin=0.22,
    ymax=0.27,
    ylabel near ticks,
    yticklabel pos=right,
    xlabel near ticks,
    symbolic x coords={1, 2, 5, 10, 15, 20, 100},
    xticklabels={1, 2, 5, 10, 15, 20, $\infty$},
    xtick=data,
    legend cell align=left,
    /pgfplots/enlargelimits=false,
          legend style={legend pos=north west,font=\tiny}
    ]
    \addplot[name path=one_top,color=cambridgeblue,smooth,line width=1.5pt] table [col sep=tab,x index=1,y index=2] {images/csv/yahoo_depth_npr1_avg_max_min.tsv};
    \addplot[name path=one_down,color=brightube,smooth, line width=1.5pt] table [col sep=tab,x index=1,y index=3] {images/csv/yahoo_depth_npr1_avg_max_min.tsv};
    \addplot[name path=one_down,color=bananamania,smooth, line width=1.5pt] table [col sep=tab,x index=1,y index=4] {images/csv/yahoo_depth_npr1_avg_max_min.tsv};
    \legend{Average, Maximum, Minimum}
    \end{axis}    
\path
  (current axis.south west) -- ++(-0.25in,-0.25in)
  rectangle (current axis.north east) -- ++(0.1in,0.1in);
\end{tikzpicture}
   \caption{}
    \label{fig:depth_abl}
\end{subfigure}
\caption{An example of how \framework models the item representation. Figure \ref{fig:sub1} shows the entities in the knowledge graph linked to the items. In figure \ref{fig:sub2}, the most relevant semantic features selected by \framework are reported.}
\label{fig:test}
\vspace{-1em}
\end{figure}
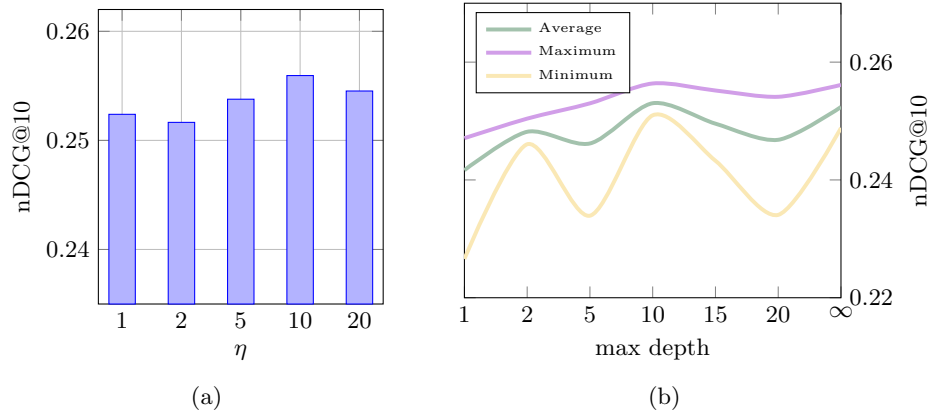

%% file: sections/conclusion.tex
\section{Conclusion and future work}
This paper addresses the need for simplification in existing KGCF recommender systems, proposing a novel algorithm called \textbf{\framework} (Knowledge Graph User-based Filtering). 
We advocate for a streamlined approach, asserting that simpler strategies have the potential to yield comparable or even superior accuracy.

In \framework, we focus on efficiently utilizing KG side information by learning user and item embeddings through linear propagation on the user-item interaction graph. Unlike other approaches, our solution integrates latent representations for semantic features directly into item representations. 
Additionally, we exploit the simple but effective decision tree mechanism to select meaningful semantic features from KGs and filter out noise triples.

We conduct extensive experiments on well-known recommendation datasets, evaluating the effectiveness of our framework. Furthermore, we explore improving the decision tree 
capability in selecting relevant semantic features. We investigate the impact of negative sampling and tree depth constraints and evaluate their impact on the overall recommendation performance.